\documentclass[aps,pra,amssymb,amsfonts,floatfix]{revtex4}

\usepackage{graphicx}
\usepackage{dcolumn}
\usepackage{bm}
\usepackage{mathbbol}

\begin{document}

\title{Manipulation of the dynamics of many-body systems via quantum control methods}

\author{Julie Dinerman}

\author{Lea F. Santos} \thanks{Corresponding author: {\tt lsantos2@yu.edu}}
\affiliation{\mbox{Department of Physics, Yeshiva University, New York, NY, 10016, USA}}

\date{\today}

\begin{abstract}
We investigate how dynamical decoupling methods may be used to manipulate the time
evolution of quantum many-body systems. These methods consist of  
sequences of external control operations designed 
to induce a desired dynamics. The systems considered for the analysis 
are one-dimensional spin-1/2 models, which, according to the parameters 
of the Hamiltonian, may be in the integrable or non-integrable limits, and in the gapped
or gapless phases. 
We show that an appropriate control sequence may lead 
a chaotic chain to evolve as an integrable chain and a system in the 
gapless phase to behave as a system in the gapped phase.
A key ingredient for the control schemes developed here is
the possibility to use, in the same sequence, 
different time intervals between control operations.
\end{abstract}

\date{\today}
\maketitle

\section{Introduction}

The study of one-dimensional quantum many-body systems out of equilibrium offers
several fundamental challenges. 
A main question is whether an isolated quantum system can or not thermalize after a quench.
While a relationship between quantum chaos and thermalization
is certain to exist \cite{Deutsch1991,Srednicki1994,Flambaum1996,ZelevinskyRep1996,rigol09,SantosPRE10},
an ongoing controversy is 
the occurrence of thermalization for a chaotic system in the gapped phase
\cite{Kollath2007,Manmana2007,Rigol2010}.
In terms of transport properties, the discussions are associated with the expected
dynamical behavior of systems in the integrable and nonintegrable regimes
\cite{Bonca1994,Zotos1999,Alvarez2002,Heidrich2003,Heidrich2004,Zotos2005,Mukerjee2008,
Steinigeweg2006,Santos2008PRE,Santos2009JMP,Santos2009JPCS}, and in the
gapped and gapless phases \cite{Langer2009,Steinigeweg2009}. Extraordinary technological 
advances have motivated the use of experiments to address these issues.
In this work we show how these experiments may benefit from dynamical decoupling methods
to accurately reach the different regimes and phases of interest.

The fundamental questions raised above have found a test-bed in 
experiments with ultracold quantum gases in optical lattices \cite{Lewenstein2007}
and with magnetic compounds \cite{Sologubenko,Hess2007,BaeriswylBook}.
Both systems require a high level of controllability. In optical lattices,
control of atomic transport is achieved with time-dependent variations 
of the depth and phase of the lattice potential;
whereas in magnetic systems the preparation of samples where impurities and unwanted
couplings are negligible is key for obtaining a desired transport behavior.
In addition to monitoring experimental conditions,
we advocate the use of sequences of external 
electromagnetic field pulses
to manipulate the dynamics of the systems.
This method of quantum control has long been employed in 
nuclear magnetic resonance (NMR) spectroscopy \cite{HaeberlenBook,ErnstBook}.
There, the Hamiltonian of nuclear spin systems is modified with sequences of 
unitary transformations (radio-frequency pulses), which constantly rotate the spins
and remove or rescale selected terms of the original Hamiltonian.
These techniques have acquired a new dimension 
in the context of quantum information 
processing \cite{Viola1998,Vandersypen2004}, 
where the studies of transport of information \cite{Cappellaro2007}
and the manipulation of electron spins with ultrafast laser pulses \cite{Press2008,Phelps2009}
are among the latest developments. Recently, these methods have  been 
proposed also for the control of atomic transport in spinor optical lattices~\cite{Mischuck2009}.

In the present work we use the theory developed in NMR to 
construct sequences of control operations that 
appropriately modify the Hamiltonian of a
one-dimensional quantum many-body system and forces it 
to follow a desired dynamics. 
The systems analyzed are modeled by spin-1/2 chains, which may also
be mapped onto systems of spinless fermions or hardcore bosons.
We focus on limits that are of interest for studies
of transport behavior and thermalization. 
Specifically, our goals are 
(i) to dynamically remove the effects
of terms leading to the onset of chaos, such as disorder and frustration,  
and compel the chain
to evolve as an integrable system, as 
introduced in Refs.~\cite{Santos2008PRE,Santos2009JMP};
(ii) to dynamically increase the effects of terms that can drive the system across
a transition from the gapless to the gapped phase (and vice-versa), aiming, for
example, at forcing a
conductor to behave as an insulator or at bringing a chaotic system into the gapped phase.
For the gapless-gapped transition, the strengths of the coupling terms in the Hamiltonian are modified by
applying to the target system 
sequences with different time intervals between the control operations.
The idea of varying the time separation between pulses has been exploited 
in sequences created to suppress the couplings between a qubit and its environment
\cite{Uhrig2007,Uhrig2009}. Contrary to that, 
the systems considered here are isolated from the environment 
and our objective is not necessarily to completely eliminate couplings, 
but instead to control their strengths.

This paper is organized as follows. Sec.~II describes the model under investigation. Sec.~III
reviews a technique to design control sequences, which is based on the average Hamiltonian theory. 
Sec.~IV presents in detail sequences that 
dynamically lead the system studied across 
a transition between the nonintegrable and integrable limits and between the
gapless and gapped phases. Numerical results are shown
for the time evolution of local magnetization 
based on a particular initial state. A discussion about 
control metrics and the dependence of the outcomes on 
initial states is left for the Appendix. Conclusions are given in Sec.~V.

\section{Description of the Model}

We study a one-dimensional spin-1/2 system with open boundary conditions described by the Hamiltonian

\begin{equation}
H_0 = H_z + H_{\text{NN}}+ \alpha H_{\text{NNN}} ,
\label{ham} 
\end{equation}
where
\begin{eqnarray}
&&H_z =  \sum_{n=1}^{L} \epsilon_n S_n^z, 
\nonumber \\
&&H_{\text{NN}} = \sum_{n=1}^{L-1} J\left[ \left( 
S_n^x S_{n+1}^x + S_n^y S_{n+1}^y \right) +
\Delta S_n^z S_{n+1}^z \right] ,
\nonumber \\
&&H_{\text{NNN}}= \sum_{n=1}^{L-2} J\left[ \left( S_n^x S_{n+2}^x + S_n^y S_{n+2}^y
\right) + \Delta S_n^z S_{n+2}^z  \right] \:.
\nonumber
\end{eqnarray}
Above,  $\hbar$ is set equal to 1,
$L$ is the number of sites, and $S^{x,y,z}_n = \sigma^{x,y,z}_n/2$ 
are the spin operators at site $n$,  $\sigma^{x,y,z}_n$ being the Pauli matrices. 

(i) The parameter $\epsilon_n$ corresponds to the Zeeman splitting of spin $n$,
as determined by a static magnetic field in the $z$ direction.
The system is clean when all $\epsilon_n$'s are equal ($\epsilon_n=\epsilon$) and it is
disordered when at least one spin $n$ has an energy splitting  different from the others
($\epsilon_n \neq \epsilon$). We therefore consider only
on-site disorder.

(ii) $H_{\text{NN}}$, also known as the XXZ Hamiltonian, corresponds to the nearest-neighbor 
(NN) exchange; $S_n^z S_{n+1}^z$ is the Ising interaction and 
$S_n^x S_{n+1}^x + S_n^y S_{n+1}^y$ is the flip-flop term.
The coupling strength $J$ and 
the anisotropy $\Delta$ are assumed positive, thus favoring antiferromagnetic order. 
By varying $\Delta$ in $H_0=H_{NN}$ we may go from 
the gapless phase ($\Delta < 1$)
to the gapped phase ($\Delta > 1$), which, at $T=0$, is followed
by the metal-insulator Mott transition \cite{Cloizeaux1966}. 
In the particular case of $\Delta=0$, we deal with 
the XY model; whereas $\Delta=1$ leads to the 
isotropic Heisenberg Hamiltonian \cite{SutherlandBook}.

(iii) The parameter
$\alpha$ refers to the ratio between the next-nearest-neighbor (NNN) 
exchange, as determined by $H_{\text{NNN}}$, and the NN couplings.
The inclusion of NNN antiferromagnetic exchange 
frustrates the chain, since NN exchange favors ferromagnetic alignment between the second neighbors.
Different phenomena are expected in the presence of frustration \cite{DiepBook}.
In the case of a closed isotropic system, for instance,
a critical value $\alpha_c\approx 0.241$ exists which separates the spin fluid phase ($\alpha<\alpha_c$) 
from the dimer phase ($\alpha>\alpha_c$) \cite{Haldane1982,Mikeska2004}.

\begin{table}[h]
\caption{Parameters of the spin-1/2 chain described by Eq.~(\ref{ham})}
\begin{center}
\begin{tabular}{|c|c|}
\hline
$\epsilon_n$ & Energy splitting of spin $n$ \\
\hline
$J$ & Coupling strength \\
\hline
$\Delta$ & Ratio between Ising interaction and flip-flop term \\
\hline
$\alpha$ & Ratio between NNN and NN couplings \\
\hline
\end{tabular}
\end{center}
\label{table:table1}
\end{table}

A clean one-dimensional spin-1/2 model 
with NN exchange only is 
integrable and solved with the Bethe Ansatz method~\cite{Bethe}.
The addition of defects~\cite{Avishai2002}, even if just one in the middle of the chain
\cite{Santos2004}, or the inclusion of
NNN couplings~\cite{Hsu1993,Kudo2005} 
may lead to the onset of quantum chaos.

\section{Design of Dynamical Decoupling Sequences}

Dynamical decoupling methods consist in the application of sequences
of unitary transformations to generate a desired effective propagator at time $t$.
The first experimental implementations of sequences of electromagnetic
pulses aiming at controlling the dynamics of molecular systems appeared in the context of
NMR spectroscopy and correspond to the
so-called spin-echo and Carr-Purcell sequences \cite{Hahn-Echo,CP-Echo}.
The basic idea of the method is to add a time-dependent control Hamiltonian,
$H_c(t)$, to the original (time-independent, in our case)
Hamiltonian of the system, $H_0$, so that the system evolution
becomes dictated by the following propagator 

\[
U(t) ={\cal T} \exp \left\{- i\int _{0}^t [ H_0 + H_c(u) ] du \right\},
\]
where ${\cal T}$ stands for time ordering. This operator may also be written
as \cite{ErnstBook,HaeberlenBook},

\[
U(t) = U_c(t) {\cal T} \exp \left\{- i\int _{0}^t [ U_c^{-1}(u) H_0 U_c(u) ] du \right\},
\]
where

\[
U_c(t) ={\cal T} \exp \left[- i\int _{0}^t  H_c(u)  du \right],
\]
depends only on the introduced perturbation $H_c(t)$ and is known as the control propagator. 
We deal with cyclic control sequences with cycle time $T_c$. This means that the perturbation 
and the control propagator are periodic, that is,
$H_c(t+nT_c)=H_c(t)$ and $U_c(t+nT_c)=U_c(t)$, $n \in \mathbb{N}$.
Since $U_c(0) =\mathbb{1} $, it follows that
the control propagator
at $T_n=nT_c$ is $U_c(nT_c) =\mathbb{1} $.
As a result, the general propagator at $T_c$ is simply
$U(T_c) = {\cal T} e^{- i\int _{0}^{T_c} [ U_c^{-1}(u) H_0 U_c(u)] du}$ 
and for $n$ cycles

\[
U(nT_c) = U(T_c)^n.
\] 
Therefore, the evolution 
of the system at any time $T_n=nT_c$ requires only the knowledge of the 
propagator after one cycle \cite{ErnstBook,HaeberlenBook}.

We consider the ideal scenario 
of arbitrarily strong and instantaneous control operations, known in NMR
as hard pulses and popularized in the field of quantum information as bang-bang control \cite{Viola1998}.
Each pulse $P_{k+1}$ is applied after a time interval of free evolution
$\tau_{k+1} =t_{k+1}-t_{k}$, where $k \in \mathbb{N}$ and $t_0=0$.
For a cyclic sequence with $m$ pulses,
the evolution operator at $T_c=m\tau$  is then given by,

\begin{eqnarray}
U(T_c)  
&=&P_m U(t_m,t_{m-1})P_{m-1} U(t_{m-1},t_{m-2}) 
\ldots P_2 U(t_2,t_1) P_1 U(t_1,0) \nonumber \\
&=& \underbrace{(P_m P_{m-1} \ldots P_1 )}
\underbrace{(P_{m-1} \ldots P_1 )^{\dagger} 
U(t_m,t_{m-1}) (P_{m-1} \ldots P_1 )}\ldots 
\underbrace{(P_2 P_1)^{\dagger} U(t_3,t_2)(P_2 P_1)}
\underbrace{P_1^{\dagger} U(t_2,t_1)P_1}
\underbrace{ U(t_1,0)} \:
\nonumber \\
&&\hspace{0.2 cm} U_c(T_c)=\mathbb{1}  
\hspace{1.2 cm}
e^{-i ( P_{m-1} \ldots P_1)^{\dagger} H_0 (P_{m-1} \ldots P_1 ) \tau_m }
\hspace{2.0 cm}
e^{-i (P_2 P_1)^{\dagger} H_0 (P_2 P_1) \tau_3 }
\hspace{0.4 cm}
e^{-i P_1^{\dagger} H_0 P_1 \tau_2 }
\hspace{0.2 cm}
e^{-i  H_0  \tau_1 }
\nonumber 
\\
&=& \exp \left[-i H_{m-1} \tau_m \right] \ldots
\exp \left[-i H_2 \tau_3 \right] \exp \left[-i H_1 \tau_2 \right]
\exp \left[-i H_0 \tau_1 \right] \nonumber \\
&=& \exp \left[-i \bar{H} T_c \right] .
\end{eqnarray}
Above, the notation 
$H_{m-1}=( P_{m-1} \ldots P_1)^{\dagger} H_0 (P_{m-1} \ldots P_1 )$ has been used.
The Hamiltonian $\bar{H}$ in the last line is obtained via Baker-Campbell-Hausdorff expansion
and is referred to as the average Hamiltonian \cite{ErnstBook,HaeberlenBook}.

The lowest order term  of $\bar{H}$ in $\tau$ corresponds to

\begin{equation}
\bar{H}^{(0)} =  \sum_{k=0}^{m-1} \frac{\tau_{k+1}}{T_c} H_k .
\label{Hk}
\end{equation}
In designing a dynamical decoupling sequence, the primary goal is to guarantee the proximity 
of $\bar{H}^{(0)}$ to the desired Hamiltonian. This is the focus of the current 
work. We note, however, that various strategies exist to eliminate some of the
higher order terms \cite{HaeberlenBook,ErnstBook}
or to reduce the accumulation of errors arising from imperfect averaging \cite{Viola2005,Khodjasteh2005,Santos2008,Rego2009}.

\section{Modifying the System Evolution}

In this section we explore how dynamical decoupling schemes
may induce transitions between different regimes and phases of a quantum many-body system.
We aim at dynamically reshaping the original Hamiltonian (\ref{ham}) to make it
as close as possible to a desired Hamiltonian $H_w$ (the subscript "$w$" is used whenever
we refer to parameters and quantities associated with the wanted evolution).
Our numerical results are performend in the full Hilbert space of dimension $D=2^L$.
We cannot take advantage of the symmetries of $H_0$ (\ref{ham}), such as
conservation of total spin in the $z$ direction, to reduce this dimension, 
because the intervals of free 
evolution of the pulsed system involve Hamiltonians [$H_k$ in Eq.~(\ref{Hk})]
that are different from the
original $H_0$.

\subsection{Non-integrable vs integrable systems}

To start, we discuss sequences of control pulses that average out the 
terms in Hamiltonian (\ref{ham})
that may lead to the onset of chaos, namely on-site disorder and NNN exchange. 
 
\subsubsection{Annihilating the effects of 
on-site disorder: $H_z+H_{NN} \rightarrow H_{NN}$ }

The first and simplest sequence that we present 
was proposed in \cite{Santos2008PRE,Santos2009JMP}
and it aims at eliminating the effects of on-site disorder. Consider a
disordered system described by $H_z + H_{NN}$ with
one or more sites having $\epsilon_n \neq \epsilon$. In order to recover the 
transport behavior of a clean chain given by $H_{NN}$, we apply, after every $\tau$, 
control operations $P$, which rotate
all spins by 180$^o$ ($\pi$-pulses) around a direction perpendicular to $z$.
Around $x$, for instance, the pulses are

\begin{equation}
P_x = \exp \left( -i \pi \sum_{n=1}^{L} S_n^x \right) = \exp(-i \pi S^x).
\end{equation}
The propagator at cycle time $T_c=2\tau$ becomes

\begin{eqnarray}
U(T_c) &=& P_x \exp \left[-i \left(H_z + H_{NN}\right) \tau\right] 
P_x \exp \left[-i \left(H_z + H_{NN}\right) \tau\right]  \nonumber \\
&=& (-1) \exp \left[-i \left(-H_z + H_{NN}\right) \tau\right]
\exp \left[-i \left(H_z + H_{NN}\right) \tau\right] ,
\nonumber
\end{eqnarray}
which leads to the following dominant term of the average Hamiltonian 

\[
\bar{H}^{(0)} = H_{NN},
\]
as desired.

\subsubsection{Annihilating the effects of NNN exchange: $H_{NN}+H_{NNN} \rightarrow H_{NN}$}

A sequence to eliminate the effects of NNN exchange (and also on-site disorder) was
introduced in \cite{Santos2008PRE} and is further investigated here. It has
eight $\pi$-pulses equally separated 
($T_c=8\tau$) and applied to selected spins, which therefore
assumes site addressability.
The pulses are

\begin{eqnarray}
P_1 &=& P_3= \prod_{k=0}^{\left\lfloor (L-1)/4 \right\rfloor}  e^{-i\pi S^x_{1+4k}}
\prod_{k=0}^{\left\lfloor (L-2)/4 \right\rfloor} e^{-i\pi S^x_{2+4k}}, \nonumber \\
P_2 &=& P_4= \prod_{k=0}^{\left\lfloor (L-3)/4 \right\rfloor} e^{-i\pi S^y_{3+4k}}
\prod_{k=0}^{\left\lfloor (L-4)/4 \right\rfloor} e^{-i\pi S^y_{4+4k}}, \nonumber \\
P_5 &=& P_7= \prod_{k=0}^{\left\lfloor (L-2)/4 \right\rfloor} e^{-i\pi S^x_{2+4k}}
\prod_{k=0}^{\left\lfloor (L-3)/4 \right\rfloor} e^{-i\pi S^x_{3+4k}}, \nonumber \\
P_6 &=& P_8= \prod_{k=0}^{\left\lfloor (L-1)/4 \right\rfloor} e^{-i\pi S^y_{1+4k}}
\prod_{k=0}^{\left\lfloor (L-4)/4 \right\rfloor} e^{-i\pi S^y_{4+4k}} .
\label{pulses}
\end{eqnarray}
They change the signs of the coupling terms in each interval of free
evolution according to Table \ref{table:8pulses}.

\begin{table}[h]
\caption{Sign changes of the coupling terms in $H_{NN}$ and $H_{NNN}$ according to the pulses
(\ref{pulses}).}
\begin{center}
\begin{tabular}{|c|c|c|c|c|c|c|c|c|}
\hline \hline 
NN couplings & $(0,\tau)$  & $(\tau,2\tau)$  & $(2\tau , 3\tau)$ &$(3\tau , 4\tau)$  & $(4\tau , 5\tau)$  & $(5\tau , 6\tau)$  & $(6\tau , 7\tau)$  & $(7\tau , 8\tau)$  \\
\hline
$S_{2k+1}^x S_{2k+2}^x$ &+ &+ &+ &+ &+ &+ &$-$ &$-$ \\
\hline
$S_{2k+2}^x S_{2k+3}^x$ &+ &+ &$-$ &$-$ &+ &+ &+ &+ \\
\hline
$S_{2k+1}^y S_{2k+2}^y$ & +&+ &+ &+ &+ &$-$ &$-$ &+ \\
\hline
$S_{2k+2}^y S_{2k+3}^y$ &+ &$-$ &$-$ &+ &+ &+ &+ &+ \\
\hline
$S_{2k+1}^z S_{2k+2}^z$ & +&+ &+ &+ &+ &$-$ &+ &$-$ \\
\hline
$S_{2k+2}^z S_{2k+3}^z$ &+ &$-$ &+ &$-$ &+ &+ &+ &+ \\
\hline \hline
NNN couplings & $(0,\tau)$  & $(\tau,2\tau)$  & $(2\tau , 3\tau)$ &$(3\tau , 4\tau)$  & $(4\tau , 5\tau)$  & $(5\tau , 6\tau)$  & $(6\tau , 7\tau)$  & $(7\tau , 8\tau)$  \\
\hline
$S_{k+1}^x S_{k+3}^x$ &+ &+& $-$& $-$&+ & +& $-$& $-$ \\
\hline
$S_{k+1}^y S_{k+3}^y$ & +&$-$&$-$& +& +& $-$& $-$& + \\
\hline
$S_{k+1}^z S_{k+3}^z$ & +&$-$& +& $-$& +& $-$& +& $-$ \\
\hline
\end{tabular}
\end{center}
\label{table:8pulses}
\end{table}
The sequence removes NNN couplings, but at the price of also reducing the effects
of NN couplings. We achieve the following dominant term for the average Hamiltonian,
\[
\bar{H}^{(0)} = \frac{H_{NN}}{2}.
\]

\begin{figure}[htb]
\vskip 0.2cm
\includegraphics[width=4.0in]{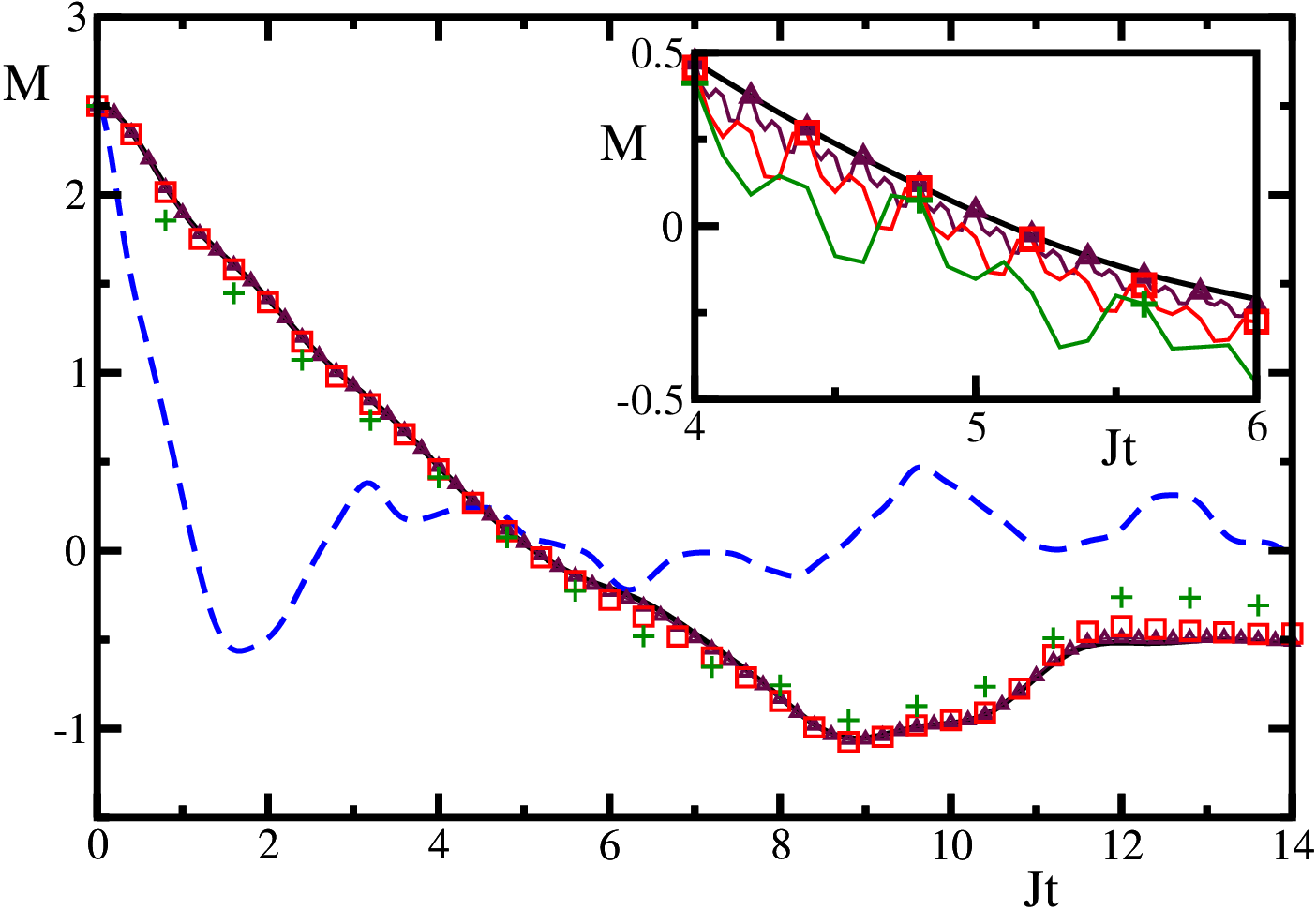}
\caption{Time evolution of local magnetization 
for a clean spin-1/2 chain with $\Delta=J/2$ and $L=10$. 
Blue dashed line: chaotic system $H_{NN}+H_{NNN}$ in the absence of pulses; 
black solid line: integrable system $H_{NN}/2$;
symbols: chaotic system $H_{NN}+H_{NNN}$ in the presence of pulses (\ref{pulses});
triangles:  $\tau=0.025 J^{-1}$, squares: $\tau=0.05 J^{-1}$,
pluses: $\tau=0.1 J^{-1}$. Main panel:
data acquired every $T_n=nT_c=8\tau$, inset: data acquired every $t_n=n\tau$.}
\label{fig1}
\end{figure}

In Fig.~\ref{fig1}, we show the evolution of a 
chaotic chain described by $H_{NN}+H_{NNN}$
under free evolution and subjected to the pulses (\ref{pulses}), 
and compare it with
the dynamics of an ideal integrable system given by $H_{NN}/2$. 
The quantity considered is
the local magnetization, which is defined as the magnetization of the first half of the chain,

\begin{equation}
M (t) \equiv \langle \Psi(t) | \sum_{n=1}^{L/2} S_{n}^z| \Psi(t) \rangle.
\label{mag}
\end{equation}
We choose as initial state $|\Psi(0)\rangle$  a
highly excited state far from equilibrium: it consists of
spins pointing up in the first 
half of the chain and  pointing down in the other half, so that $M(0)=L/4$. A
discussion about control metrics and the dependence of the results on different initial states is
given in the Appendix. 

In order to average out unwanted terms of the Hamiltonian, we need to 
consider cycle times smaller than the time scale of the free evolution of the 
system, that is, $JT_c<1$ \cite{noteCONVERGE}. For the $T_c$'s considered
in Fig.~\ref{fig1},
the agreement between the evolutions of the
chaotic system under control pulses and the integrable chain is excellent.
As the values of $\tau$ increase, higher order terms in the average Hamiltonian 
become significant, specially at long times, and
the pulsed system must eventually separate from the ideal scenario. 
Even though strategies exist to reduce the
effects of higher order terms in $\bar{H}$
\cite{Viola2005,Khodjasteh2005,Santos2008,Rego2009}, the access to short time intervals between
pulses is essential to the success of dynamical decoupling methods.

Note that control sequences are designed to match the aspired evolution at the completion
of cycles, but not necessarily in between cycles. 
This is well illustrated in the inset.
The symbols
indicate data obtained at every $T_n=nT_c$, while the solid lines
correspond to the values of $M(t)$ at every $t_n=n\tau$. 
The largest mismatches between the evolution determined by $H_{NN}+H_{NNN}+H_c(t)$
and by $H_{NN}/2$ occur in between cycles.  

A second alternative to recover the dynamical behavior of an integrable system
consists in canceling out the NN exchange leaving the NNN couplings unaffected.
This is achieved with a sequence of four pulses equally separated ($T_c=4\tau$), 
which alternates rotations around perpendicular directions, say $x$ and $y$.
The pulses can affect only spins in odd sites, or only in even sites, or they alternate
between odd and even sites as below,

\begin{equation}
P_1 =P_3= \prod_{k=0}^{\left\lfloor (L-1)/2 \right\rfloor} e^{-i\pi S^y_{1+2k}},
\hspace{0.8 cm}
P_2 = P_4= \prod_{k=0}^{\left\lfloor(L-2)/2\right\rfloor}  e^{-i\pi S^x_{2+2k}}.
\label{pulses_NN}
\end{equation}
The pulses (\ref{pulses_NN}) change the signs of the terms of 
$H_{NN}+H_{NNN}$ as shown in Table \ref{table:4pulses}.

\begin{table}[h]
\caption{Sign changes of the coupling terms in $H_{NN}$ and $H_{NNN}$ according to the pulses
(\ref{pulses_NN}).}
\begin{center}
\begin{tabular}{|c|c|c|c|c|}
\hline \hline 
NN couplings & $(0,\tau)$  & $(\tau,2\tau)$  & $(2\tau , 3\tau)$ &$(3\tau , 4\tau)$   \\
\hline
$S_{k+1}^x S_{k+2}^x$ &+ &$-$ &$-$ &+ \\
\hline
$S_{k+1}^y S_{k+2}^y$ & +&+ &$-$ &$-$  \\
\hline
$S_{k+1}^z S_{k+2}^z$ & +&$-$ &+ &$-$  \\
\hline \hline
NNN couplings & $(0,\tau)$  & $(\tau,2\tau)$  & $(2\tau , 3\tau)$ &$(3\tau , 4\tau)$   \\
\hline
$S_{k+1}^x S_{k+3}^x$ &+ &+& +& + \\
\hline
$S_{k+1}^y S_{k+3}^y$ & +&+&+& + \\
\hline
$S_{k+1}^z S_{k+3}^z$ & +&+& +& +\\
\hline
\end{tabular}
\end{center}
\label{table:4pulses}
\end{table}
The above control sequence leads to two uncoupled chains, one consisting of 
exchange couplings between spins in 
the original odd sites and the other one consisting of 
couplings between spins in the original even sites. The first order term in the average
Hamiltonian is simply,

\[
\bar{H}^{(0)} = H_{NNN}.
\]


\subsection{Gapped vs gapless systems}

Here, we analyze how the time evolution typical of a system with an energy gap may be achieved from a gapless system subjected to a quantum control scheme. Instead of 
completely averaging out unwanted terms of the Hamiltonian,
as done in the previous section, the goal now is to change the ratio 
between different coupling strengths.
This is accomplished by varying the time interval between pulses,
so that the weight of some specific terms may be decreased relatively to others.

\subsubsection{Increasing the role of the Ising interaction: 
$\Delta <1 \rightarrow \Delta >1$}

Consider a given system in the gapless phase described by $H_{NN}$, where 
$\Delta<1$. To induce the transport behavior of a chain in the gapped phase, 
we develop a 
sequence that leaves the Ising interaction unaffected, but reduces the strength of the 
flip-flop term. The sequence consists of two $\pi$-rotations around the $z$ axis, 
which affect only spins in even sites (or only in odd sites), that is,

\begin{equation}
P_1=P_2 = \exp \left( -i \pi \sum_{k=0}^{\left\lfloor (L-2)/2 \right\rfloor} 
S_{2k+2}^z \right) .
\label{pulses_gap}
\end{equation}
The first pulse is applied after $\tau_1$ and the second after $\tau_2$, which results in the following
reshaped Hamiltonian at $T_c=\tau_1+\tau_2$,

\[
\bar{H}^{(0)} = \sum_{n=1}^{L-1} J\left[\frac{(\tau_1-\tau_2)}{T_c} \left(
S_n^x S_{n+1}^x + S_n^y S_{n+1}^y \right)+ 
\Delta S_n^z S_{n+1}^z \right].
\]
The gapped phase is obtained if $ T_c\Delta/(\tau_1-\tau_2) >1$.

\begin{figure}[htb]
\vskip 0.2cm
\includegraphics[width=4.5in]{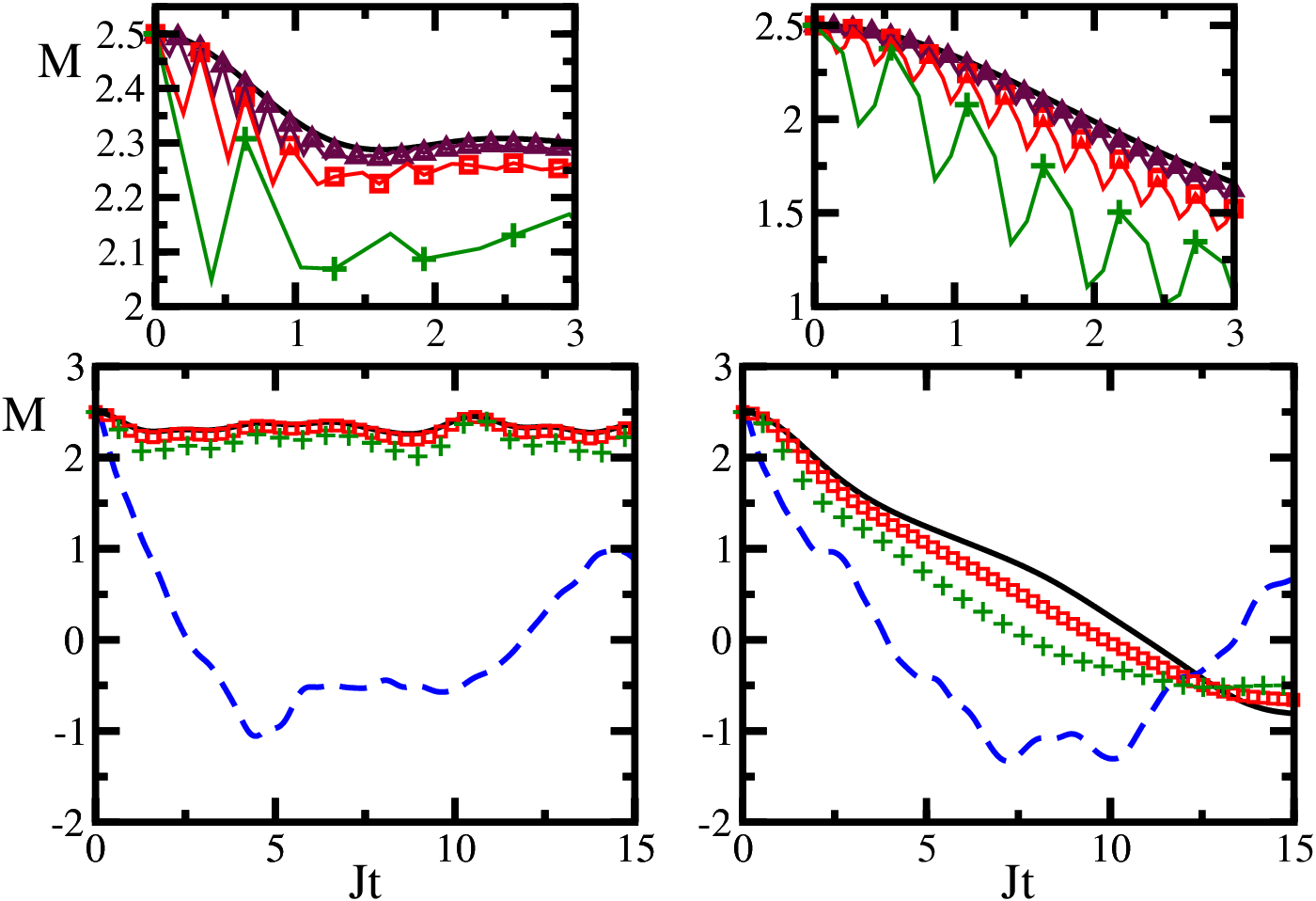}
\caption{Evolution of local magnetization for a clean spin-1/2 chain with $L=10$, 
described by $H_{NN}$ 
with $\Delta=1/2$ (left panels) and 
by $H_{NN}+\alpha H_{NNN}$ with
$\Delta =1$ and $\alpha=0.1$ (right panels).
Absence of pulses: blue dashed line; presence
of pulses: symbols. 
Left panels:
pulses (\ref{pulses_gap}) leading to $T_c\Delta/(\tau_1-\tau_2)=2$.
Triangles (appearing only in the inset): $\tau_1=0.1 J^{-1}$ and 
$\tau_2=0.06 J^{-1}$; squares: $\tau_1=0.2 J^{-1}$ and 
$\tau_2=0.12  J^{-1}$; pluses: $\tau_1=0.4 J^{-1}$ and 
$\tau_2=0.24  J^{-1}$.
Black solid line: 
system described by $H_{NN}/4$ with $\Delta_w=2$.
Right panels: pulses (\ref{pulses_NN}) 
leading to 
$\alpha T_c/(\tau_1-\tau_a)=0.64$, where
$\tau_a=\tau_{2,3,4}=\tau_1 (0.64-\alpha)/(0.64+3\alpha)$.
Triangles (appearing only in the inset): $\tau_1=0.05 J^{-1}$;
squares: $\tau_1=0.1 J^{-1}$; pluses: $\tau_1=0.2 J^{-1}$.
Black solid line: system described by 
$(H_{NN}+ \alpha_w H_{NNN})/6.4$ with $\Delta=1$ and $\alpha_w=0.64$.
Bottom panels: data acquired every cycle; top panels: data acquired
every pulse.
}
\label{fig2}
\end{figure}

On the left panels of Fig.~\ref{fig2} we compare 
the desired evolution of a gapped system described by $H_{NN}/4$, where $\Delta_w=2$,
with the dynamics of a given gapless system described by
$H_{NN}$, where $\Delta=1/2$, and subjected to control operations.
The pulses, given by Eq.~(\ref{pulses_gap}), have time intervals 
that lead to $T_c \Delta /(\tau_1-\tau_2)=2$. The agreement 
between the two cases at the completion of
each cycle is very good for the chosen time intervals, but of course
it worsens as $T_c$ increases. The price to dynamically achieve a transition from 
a gapless to a gapped Hamiltonian, however, is the slowing down of the 
system evolution in the gapped phase by a factor of 4. For 
a discussion about ballistic vs diffusive transport in XXZ models with small and 
large values of $\Delta$, see Refs.~\cite{Langer2009,Steinigeweg2009}
and references therein.

It is also possible to force a transition from a gapped to a gapless phase.
For example, pulses (\ref{pulses_NN}) separated by different intervals
$\tau_1$, $\tau_2=\tau_4$, and $\tau_3$, reshape 
the Hamiltonian of a Mott insulator described by $H_{NN}$ with $\Delta>1$ as
follows,

\[
\bar{H}^{(0)} =  \sum_{n=1}^{L-1} \frac{J}{T_c}\left[  (\tau_1-\tau_3)
\left(S_n^x S_{n+1}^x + S_n^y S_{n+1}^y \right) +
\Delta (\tau_1+\tau_3 - 2\tau_2) S_n^z S_{n+1}^z \right].
\]
The gapless phase is then obtained if $\Delta (\tau_1+\tau_3 - 2\tau_2)< (\tau_1-\tau_3)$.
In particular, when $\tau_1+\tau_3 = 2\tau_2$ we recover the XY model.

\subsubsection{From the spin fluid to the dimer phase: 
$\alpha <\alpha_c \rightarrow \alpha >\alpha_c$}

Let us now focus on a system described by 
$H_{NN}+\alpha H_{NNN}$ with
$\alpha<\alpha_c$ and $\Delta=1$. 
Our goal is to apply to it a sequence of control pulses that induces 
a dynamical behavior similar to that of a system in the dimerized phase ($\alpha >\alpha_c$). 
A possible sequence consists of the four pulses presented in Eq.~(\ref{pulses_NN}), although now
separated by different time intervals 
$\tau_1$ and $\tau_2 = \tau_3 =\tau_4 = \tau_a$. 
It leads to the following dominant term of the average Hamiltonian at 
$T_c = \tau_1+\tau_2 + \tau_3 +\tau_4$,

\[
\bar{H}^{(0)} = \frac{(\tau_1-\tau_a)}{T_c} H_{NN} + \alpha H_{NNN}.
\]
The desired dynamics is reached when $\alpha T_c/(\tau_1-\tau_a) > \alpha_c$.

On the right panels of Fig.~\ref{fig2}, we compare the dynamics of a 
system in the gapless phase given by 
$H_{NN}+\alpha H_{NNN}$, where  $\alpha=0.1$ and $\Delta=1$, with the wanted
dynamics of a system in the gapped phase described by
$(H_{NN} + \alpha H_{NNN})/6.4$, where $\alpha_w=0.64$ and $\Delta=1$. The 
pulses~(\ref{pulses_NN}) applied to the original system
are separated by intervals that satisfy the constraint $\alpha T_c/(\tau_1-\tau_a)=0.64$. 
For $JT_c<1$, the dynamics of the perturbed gapless system at the completion of 
the cycles agrees well
with the evolution of the gapped system.

Notice that our pulsed system will now be in the limit of strong
frustration and will therefore become chaotic \cite{Kudo2005}.
We expect an initial state with energy far from the edges of the spectrum 
to show a dynamical behavior typical of a chaotic system 
\cite{noteCHAOS}, although slowed down by a factor $\alpha_w/\alpha$.

A variety of new pulse sequences may be designed according to our needs and 
experimental capabilities. For instance, the sequence described above, 
which has no effect on NNN couplings, could be combined
with a sequence of $\pi$-pulses in the $z$ direction given by

\[
P_z = \prod_{k=0}^{\left\lfloor (L-1)/4 \right\rfloor}  e^{-i\pi S^z_{1+4k}}
\prod_{k=0}^{\left\lfloor (L-2)/4 \right\rfloor} e^{-i\pi S^z_{2+4k}}.
\]
These pulses would affect only the flip-flop term of $H_{NNN}$. The combined 
sequence would reduce the strength of $H_{NN}$ and keep only the Ising 
interaction from $H_{NNN}$. Hamiltonians
similar to this resulting one have been analyzed in the 
context of transport \cite{Rabson2004} and thermalization in gapped systems \cite{Rigol2010}.

\section{Conclusion}

We studied how quantum control methods may be used to manipulate the dynamics of 
quantum many-body systems. This is accomplished with sequences of unitary transformations designed to achieve a desired evolution operator. The unitary 
transformations correspond to instantaneous
electromagnetic field pulses, which, in the case of spin systems, rotate the
spins by a certain angle. The strategy employed to construct the control sequences
was the average Hamiltonian theory.

Four target spin-1/2 chains were considered. Two were chaotic:
(i) disordered with NN exchange, (ii) clean with NN exchange 
and NNN frustrating couplings;
and two were in the gapless phase: (iii) 
clean with NN couplings and small Ising interaction,
(iv) clean with NN couplings and weak frustration. 
We discussed sequences of control pulses aiming at averaging out the effects of disorder, in case (i), 
and the effects of NNN couplings, in case (ii), and then induce in these chaotic systems a 
dynamical behavior
typical of an integrable clean chain with only NN exchange. 
We also developed new sequences capable of changing the ratio between the coupling strengths of the system. This was done by exploiting, 
in the same sequence, different time intervals between the pulses,
so that the effects of specific terms in the Hamiltonian may be reduced, but not 
necessarily eliminated.
By dynamically decreasing the weight of the flip-flop term in (iii), we obtained a time
evolution characteristic of gapped systems with large Ising interaction 
(also mentioned was a scheme to achieve the dynamics of a metal out of an insulator).
By decreasing the strength of the NN couplings with respect to the 
NNN exchange in (iv), 
we obtained the dynamical behavior of a chain in the limit of strong frustration
(also mentioned was a way to combine to this scheme a sequence that removes the NNN flip-flop term).

Site addressability and the possibility to vary the time intervals between 
pulses add versatility to dynamical decoupling schemes and open an array 
of new opportunities for controlling quantum systems. Our focus here was to use
these ideas as an additional tool to reach regimes and phases
that are relevant in studies of the nonequilibrium dynamics of many-body systems.

\begin{acknowledgments}
J.D. thanks financial aid from the S. Daniel Abraham Honors Program at Stern College for Women.
This work was supported by a grant from the Research Corporation.
\end{acknowledgments}

\appendix

\section{Control Metric}

In the main text, we showed numerical results for the local magnetization.
This choice was motivated by works on transport properties and
thermalization, where quantities of experimental interest are usually
considered. The decision to take the particular initial state 
where all up-spins are in the first half of the chain
was a way 
to connect our results with previous transport studies \cite{Steinigeweg2006,Santos2008PRE,Santos2009JMP},
and also to provide a good illustration,
since states where $M(0) \sim 0$ would not give much information.
However, in order to quantify how successful the designed control sequences 
are in general, we need an appropriate control metric which is independent of the
initial state.

How close a pulse sequence brings the evolution of a system to a
desired dynamics 
given by $U_{w}(t)$ may be assessed
by the so-called propagator fidelity, which is defined as
\cite{Bowdrey2002},

\begin{equation}
F_u (t) \equiv | Tr [U_{w}^{\dagger}(t) U(t)]|/D.
\label{fidel}
\end{equation}
Above, $D=2^L$ is the dimension of the system 
and $U(t)$ dictates
the evolution of the pulsed system. This quantity deals
directly with propagators and therefore avoid any mention to 
particular initial states.
Our goal is to achieve $F_u (t) \rightarrow 1$.

\begin{figure}[htb]
\vskip 0.2cm
\includegraphics[width=4.5in]{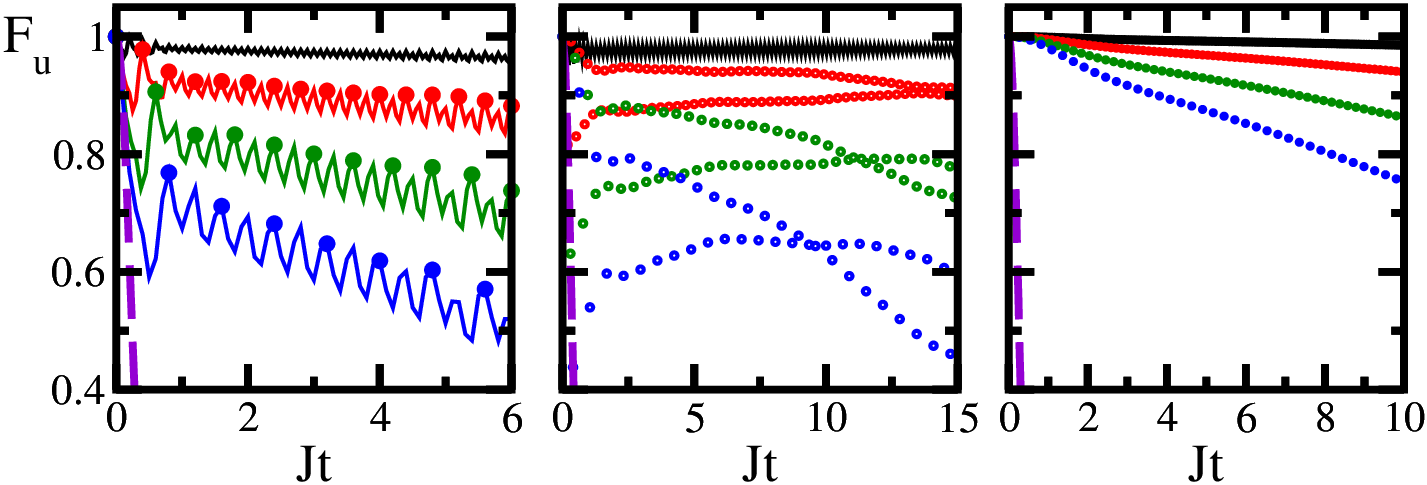}
\caption{Propagator fidelity vs time for clean spin-1/2 chains,  $L=10$.
Left panel: propagator for  $H_{w} = H_{NN}/2$ vs propagator for a
pulsed system described by $H_{NN}+H_{NNN}$;
$\Delta=1/2$ for both. From top to bottom: pulses (\ref{pulses})
separated by $\tau=0.025 J^{-1}, 0.05 J^{-1}, 0.075 J^{-1}, 0.1 J^{-1}$.
Data acquired after every pulse, cycles indicated with symbols.
Middle panel: propagator for $H_{w}=H_{NN}/4$ with $\Delta=2$ 
vs
propagator for pulsed $H_{NN}$ with $\Delta=1/2$.
Data acquired after every pulse and indicated with symbols. From top
to bottom: pulses (\ref{pulses_gap})
separated by $\tau_1=0.1 J^{-1}$ and $\tau_2=0.06 J^{-1}$; 
$\tau_1=0.2 J^{-1}$ and $\tau_2=0.12 J^{-1}$;
$\tau_1=0.3 J^{-1}$ and $\tau_2=0.18 J^{-1}$;
$\tau_1=0.4 J^{-1}$ and $\tau_2=0.24 J^{-1}$.
Right panel: propagator for 
$H_{w}=(H_{NN}+\alpha_w H_{NNN})/6.4$ 
with  $\alpha_w=0.64$
vs propagator
for pulsed $H_{NN}+\alpha H_{NNN}$ with
 $\alpha=0.1$; $\Delta=1$ for both.
From top to bottom: pulses (\ref{pulses_NN})
separated by $\tau_1=0.025 J^{-1}, 0.05 J^{-1},
0.075 J^{-1}, 0.1 J^{-1}$
and
$\tau_a=\tau_{2,3,4}=\tau_1 (0.64-\alpha)/(0.64+3\alpha)$.
Data acquired after every cycle.
All panels: fast decaying dashed line -- absence of pulses.
}
\label{fig3}
\end{figure}

In Fig.~\ref{fig3}, we show again the cases studied in 
Figs.~\ref{fig1} and \ref{fig2}, but now from the perspective
of the evolution operator. The performance of the pulse sequences deteriorates
for longer $T_c$'s and longer times.
For the shortest cycle times considered, however,
the agreement between the pulsed and desired dynamics is very good for 
relatively long times, giving a proof of principle that indeed
we may manipulate the dynamics of quantum many-body systems
with dynamical decoupling methods 
\cite{noteBEST}.

The agreement between $U_w(t)$ and $U(t)$ is best at the completion of cycles, as shown in the 
left panel and also as verified, but not shown, for the system in the right panel.
Interestingly, however, for the system in the middle panel this 
does not always hold.
In the middle panel we compare the gapless system  $H_{NN}$ with $\Delta=1/2$
and the gapped system 
$H_{w}=H_{NN}/4$ with $\Delta=2$. The applied pulses
involve a single direction $z$ and the cycle is 
completed after only two pulses [cf. Eq.~(\ref{pulses_gap})]. In the figure, we show with circles the
values of $F_u$ after every pulse, so each value of the pair
$\tau_{1,2}$ has associated with it two curves. For short times, the performance of the sequence is best
at the completion of the cycles, as expected, but at longer times, the scenario is reversed
and the best performance appears in between cycles. The origin of this shift of roles,
which occurs faster for longer cycle times, deserves further investigation.


\begin{thebibliography}{58}
\expandafter\ifx\csname natexlab\endcsname\relax\def\natexlab#1{#1}\fi
\expandafter\ifx\csname bibnamefont\endcsname\relax
  \def\bibnamefont#1{#1}\fi
\expandafter\ifx\csname bibfnamefont\endcsname\relax
  \def\bibfnamefont#1{#1}\fi
\expandafter\ifx\csname citenamefont\endcsname\relax
  \def\citenamefont#1{#1}\fi
\expandafter\ifx\csname url\endcsname\relax
  \def\url#1{\texttt{#1}}\fi
\expandafter\ifx\csname urlprefix\endcsname\relax\def\urlprefix{URL }\fi
\providecommand{\bibinfo}[2]{#2}
\providecommand{\eprint}[2][]{\url{#2}}

\bibitem[{\citenamefont{Deutsch}(1991)}]{Deutsch1991}
\bibinfo{author}{\bibfnamefont{J.~M.} \bibnamefont{Deutsch}},
  \bibinfo{journal}{Phys. Rev. A} \textbf{\bibinfo{volume}{43}},
  \bibinfo{pages}{2046} (\bibinfo{year}{1991}).

\bibitem[{\citenamefont{Srednicki}(1994)}]{Srednicki1994}
\bibinfo{author}{\bibfnamefont{M.}~\bibnamefont{Srednicki}},
  \bibinfo{journal}{Phys. Rev. E} \textbf{\bibinfo{volume}{50}},
  \bibinfo{pages}{888} (\bibinfo{year}{1994}).

\bibitem[{\citenamefont{Flambaum et~al.}(1996)\citenamefont{Flambaum, Izrailev,
  and Casati}}]{Flambaum1996}
\bibinfo{author}{\bibfnamefont{V.~V.} \bibnamefont{Flambaum}},
  \bibinfo{author}{\bibfnamefont{F.~M.} \bibnamefont{Izrailev}},
  \bibnamefont{and} \bibinfo{author}{\bibfnamefont{G.}~\bibnamefont{Casati}},
  \bibinfo{journal}{Phys. Rev. E} \textbf{\bibinfo{volume}{54}},
  \bibinfo{pages}{2136} (\bibinfo{year}{1996}).

\bibitem[{\citenamefont{Zelevinsky et~al.}(1996)\citenamefont{Zelevinsky,
  Brown, Frazier, and Horoi}}]{ZelevinskyRep1996}
\bibinfo{author}{\bibfnamefont{V.}~\bibnamefont{Zelevinsky}},
  \bibinfo{author}{\bibfnamefont{B.~A.} \bibnamefont{Brown}},
  \bibinfo{author}{\bibfnamefont{N.}~\bibnamefont{Frazier}}, \bibnamefont{and}
  \bibinfo{author}{\bibfnamefont{M.}~\bibnamefont{Horoi}},
  \bibinfo{journal}{Phys. Rep.} \textbf{\bibinfo{volume}{276}},
  \bibinfo{pages}{85} (\bibinfo{year}{1996}).

\bibitem[{rig()}]{rigol09}
\bibinfo{note}{M. Rigol and V. Dunjko and M. Olshanii, Nature {\bf 452}, 854
  (2008); M. Rigol, Phys. Rev. Lett. {\bf 103}, 100403 (2009); M. Rigol,
  arXiv:0908.3188.}

\bibitem[{\citenamefont{Santos and Rigol}(2010)}]{SantosPRE10}
\bibinfo{author}{\bibfnamefont{L.~F.} \bibnamefont{Santos}} \bibnamefont{and}
  \bibinfo{author}{\bibfnamefont{M.}~\bibnamefont{Rigol}},
  \bibinfo{journal}{Phys. Rev. E} \textbf{\bibinfo{volume}{81}},
  \bibinfo{pages}{036206} (\bibinfo{year}{2010}).

\bibitem[{\citenamefont{Kollath et~al.}(2007)\citenamefont{Kollath,
  L{\"a}uchli, and Altman}}]{Kollath2007}
\bibinfo{author}{\bibfnamefont{C.}~\bibnamefont{Kollath}},
  \bibinfo{author}{\bibfnamefont{A.~M.} \bibnamefont{L{\"a}uchli}},
  \bibnamefont{and} \bibinfo{author}{\bibfnamefont{E.}~\bibnamefont{Altman}},
  \bibinfo{journal}{Phys. Rev. Lett.} \textbf{\bibinfo{volume}{98}},
  \bibinfo{pages}{180601} (\bibinfo{year}{2007}).

\bibitem[{\citenamefont{Manmana et~al.}(2007)\citenamefont{Manmana, Wessel,
  Noack, and Muramatsu}}]{Manmana2007}
\bibinfo{author}{\bibfnamefont{S.~R.} \bibnamefont{Manmana}},
  \bibinfo{author}{\bibfnamefont{S.}~\bibnamefont{Wessel}},
  \bibinfo{author}{\bibfnamefont{R.~M.} \bibnamefont{Noack}}, \bibnamefont{and}
  \bibinfo{author}{\bibfnamefont{A.}~\bibnamefont{Muramatsu}},
  \bibinfo{journal}{Phys. Rev. Lett.} \textbf{\bibinfo{volume}{98}},
  \bibinfo{pages}{210405} (\bibinfo{year}{2007}).

\bibitem[{\citenamefont{Rigol and Santos}()}]{Rigol2010}
\bibinfo{author}{\bibfnamefont{M.}~\bibnamefont{Rigol}} \bibnamefont{and}
  \bibinfo{author}{\bibfnamefont{L.~F.} \bibnamefont{Santos}},
  \bibinfo{note}{arXiv:1003.1403}.

\bibitem[{\citenamefont{Bonca et~al.}(1994)\citenamefont{Bonca, Rodriguez,
  Ferrer, and Bedell}}]{Bonca1994}
\bibinfo{author}{\bibfnamefont{J.}~\bibnamefont{Bonca}},
  \bibinfo{author}{\bibfnamefont{J.~P.} \bibnamefont{Rodriguez}},
  \bibinfo{author}{\bibfnamefont{J.}~\bibnamefont{Ferrer}}, \bibnamefont{and}
  \bibinfo{author}{\bibfnamefont{K.~S.} \bibnamefont{Bedell}},
  \bibinfo{journal}{Phys. Rev. B} \textbf{\bibinfo{volume}{50}},
  \bibinfo{pages}{3415} (\bibinfo{year}{1994}).

\bibitem[{\citenamefont{Zotos}(1999)}]{Zotos1999}
\bibinfo{author}{\bibfnamefont{X.}~\bibnamefont{Zotos}},
  \bibinfo{journal}{Phys. Rev. Lett.} \textbf{\bibinfo{volume}{82}},
  \bibinfo{pages}{1764} (\bibinfo{year}{1999}).

\bibitem[{\citenamefont{Alvarez and Gros}(2002)}]{Alvarez2002}
\bibinfo{author}{\bibfnamefont{J.~V.} \bibnamefont{Alvarez}} \bibnamefont{and}
  \bibinfo{author}{\bibfnamefont{C.}~\bibnamefont{Gros}},
  \bibinfo{journal}{Phys. Rev. Lett.} \textbf{\bibinfo{volume}{88}},
  \bibinfo{pages}{077203} (\bibinfo{year}{2002}).

\bibitem[{\citenamefont{Heidrich-Meisner
  et~al.}(2003)\citenamefont{Heidrich-Meisner, Honecker, Cabra, and
  Brenig}}]{Heidrich2003}
\bibinfo{author}{\bibfnamefont{F.}~\bibnamefont{Heidrich-Meisner}},
  \bibinfo{author}{\bibfnamefont{A.}~\bibnamefont{Honecker}},
  \bibinfo{author}{\bibfnamefont{D.~C.} \bibnamefont{Cabra}}, \bibnamefont{and}
  \bibinfo{author}{\bibfnamefont{W.}~\bibnamefont{Brenig}},
  \bibinfo{journal}{Phys. Rev. B} \textbf{\bibinfo{volume}{68}},
  \bibinfo{pages}{134436} (\bibinfo{year}{2003}).

\bibitem[{\citenamefont{Heidrich-Meisner
  et~al.}(2004)\citenamefont{Heidrich-Meisner, Honecker, Cabra, and
  Brenig}}]{Heidrich2004}
\bibinfo{author}{\bibfnamefont{F.}~\bibnamefont{Heidrich-Meisner}},
  \bibinfo{author}{\bibfnamefont{A.}~\bibnamefont{Honecker}},
  \bibinfo{author}{\bibfnamefont{D.~C.} \bibnamefont{Cabra}}, \bibnamefont{and}
  \bibinfo{author}{\bibfnamefont{W.}~\bibnamefont{Brenig}},
  \bibinfo{journal}{Phys. Rev. Lett.} \textbf{\bibinfo{volume}{92}},
  \bibinfo{pages}{069703} (\bibinfo{year}{2004}).

\bibitem[{\citenamefont{Zotos}(2005)}]{Zotos2005}
\bibinfo{author}{\bibfnamefont{X.}~\bibnamefont{Zotos}}, \bibinfo{journal}{J.
  Phys. Soc. Jpn} \textbf{\bibinfo{volume}{74 Suppl.}}, \bibinfo{pages}{173}
  (\bibinfo{year}{2005}).

\bibitem[{\citenamefont{Mukerjee and Shastry}(2008)}]{Mukerjee2008}
\bibinfo{author}{\bibfnamefont{S.}~\bibnamefont{Mukerjee}} \bibnamefont{and}
  \bibinfo{author}{\bibfnamefont{B.~S.} \bibnamefont{Shastry}},
  \bibinfo{journal}{Phys. Rev. B} \textbf{\bibinfo{volume}{77}},
  \bibinfo{pages}{245131} (\bibinfo{year}{2008}).

\bibitem[{\citenamefont{Steinigeweg et~al.}(2006)\citenamefont{Steinigeweg,
  Gemmer, and Michel}}]{Steinigeweg2006}
\bibinfo{author}{\bibfnamefont{R.}~\bibnamefont{Steinigeweg}},
  \bibinfo{author}{\bibfnamefont{J.}~\bibnamefont{Gemmer}}, \bibnamefont{and}
  \bibinfo{author}{\bibfnamefont{M.}~\bibnamefont{Michel}},
  \bibinfo{journal}{Europhys. Lett.} \textbf{\bibinfo{volume}{75}},
  \bibinfo{pages}{406} (\bibinfo{year}{2006}).

\bibitem[{\citenamefont{Santos}(2008)}]{Santos2008PRE}
\bibinfo{author}{\bibfnamefont{L.~F.} \bibnamefont{Santos}},
  \bibinfo{journal}{Phys. Rev. E} \textbf{\bibinfo{volume}{78}},
  \bibinfo{pages}{031125} (\bibinfo{year}{2008}).

\bibitem[{\citenamefont{Santos}(2009)}]{Santos2009JMP}
\bibinfo{author}{\bibfnamefont{L.~F.} \bibnamefont{Santos}},
  \bibinfo{journal}{J. Math. Phys} \textbf{\bibinfo{volume}{50}},
  \bibinfo{pages}{095211} (\bibinfo{year}{2009}).

\bibitem[{\citenamefont{Santos}()}]{Santos2009JPCS}
\bibinfo{author}{\bibfnamefont{L.~F.} \bibnamefont{Santos}}, \bibinfo{note}{to
  appear in J. Phys.: Conf. Ser. (2010)}.

\bibitem[{\citenamefont{Langer et~al.}(2009)\citenamefont{Langer,
  Heidrich-Meisner, Gemmer, McCulloch, and Schollw\"ock}}]{Langer2009}
\bibinfo{author}{\bibfnamefont{S.}~\bibnamefont{Langer}},
  \bibinfo{author}{\bibfnamefont{F.}~\bibnamefont{Heidrich-Meisner}},
  \bibinfo{author}{\bibfnamefont{J.}~\bibnamefont{Gemmer}},
  \bibinfo{author}{\bibfnamefont{I.~P.} \bibnamefont{McCulloch}},
  \bibnamefont{and}
  \bibinfo{author}{\bibfnamefont{U.}~\bibnamefont{Schollw\"ock}},
  \bibinfo{journal}{Phys. Rev. B} \textbf{\bibinfo{volume}{79}},
  \bibinfo{pages}{214409} (\bibinfo{year}{2009}).

\bibitem[{\citenamefont{Steinigeweg and Gemmer}(2009)}]{Steinigeweg2009}
\bibinfo{author}{\bibfnamefont{R.}~\bibnamefont{Steinigeweg}} \bibnamefont{and}
  \bibinfo{author}{\bibfnamefont{J.}~\bibnamefont{Gemmer}},
  \bibinfo{journal}{Phys. Rev. B} \textbf{\bibinfo{volume}{80}},
  \bibinfo{pages}{184402} (\bibinfo{year}{2009}).

\bibitem[{\citenamefont{Lewenstein et~al.}(2007)\citenamefont{Lewenstein,
  Sanpera, Ahufinger, Damski, De, and Sen}}]{Lewenstein2007}
\bibinfo{author}{\bibfnamefont{M.}~\bibnamefont{Lewenstein}},
  \bibinfo{author}{\bibfnamefont{A.}~\bibnamefont{Sanpera}},
  \bibinfo{author}{\bibfnamefont{V.}~\bibnamefont{Ahufinger}},
  \bibinfo{author}{\bibfnamefont{B.}~\bibnamefont{Damski}},
  \bibinfo{author}{\bibfnamefont{A.~S.} \bibnamefont{De}}, \bibnamefont{and}
  \bibinfo{author}{\bibfnamefont{U.}~\bibnamefont{Sen}}, \bibinfo{journal}{Adv.
  Phys.} \textbf{\bibinfo{volume}{56}}, \bibinfo{pages}{243}
  (\bibinfo{year}{2007}).

\bibitem[{Sol()}]{Sologubenko}
\bibinfo{note}{A. V. Sologubenko and K. Giann\`o and H. R. Ott and U. Ammerahl
  and A. Revcolevschi, Phys. Rev. Lett. {\bf 84}, 2714 (2000); A. V.
  Sologubenko and K. Giann\`o and H. R. Ott and A. Vietkine and A.
  Revcolevschi, Phys. Rev. B {\bf 64}, 054412 (2001).}

\bibitem[{\citenamefont{Hess}(2007)}]{Hess2007}
\bibinfo{author}{\bibfnamefont{C.}~\bibnamefont{Hess}}, \bibinfo{journal}{Eur.
  Phys. J. Special Topics} \textbf{\bibinfo{volume}{151}}, \bibinfo{pages}{73}
  (\bibinfo{year}{2007}).

\bibitem[{\citenamefont{Baeriswyl and Degiorgi}(2004)}]{BaeriswylBook}
\bibinfo{editor}{\bibfnamefont{D.}~\bibnamefont{Baeriswyl}} \bibnamefont{and}
  \bibinfo{editor}{\bibfnamefont{L.}~\bibnamefont{Degiorgi}}, eds.,
  \emph{\bibinfo{title}{Strong Interactions in Low Dimensions}}
  (\bibinfo{publisher}{Kluwer Academic Publishers},
  \bibinfo{address}{Dordrecht}, \bibinfo{year}{2004}).

\bibitem[{\citenamefont{Haeberlen}(1976)}]{HaeberlenBook}
\bibinfo{author}{\bibfnamefont{U.}~\bibnamefont{Haeberlen}},
  \emph{\bibinfo{title}{High Resolution NMR in Solids: Selective Averaging}}
  (\bibinfo{publisher}{Academic Press}, \bibinfo{address}{New York},
  \bibinfo{year}{1976}).

\bibitem[{\citenamefont{Ernst et~al.}(2004)\citenamefont{Ernst, Bodenhausen,
  and Wokaun}}]{ErnstBook}
\bibinfo{author}{\bibfnamefont{R.~R.} \bibnamefont{Ernst}},
  \bibinfo{author}{\bibfnamefont{G.}~\bibnamefont{Bodenhausen}},
  \bibnamefont{and} \bibinfo{author}{\bibfnamefont{A.}~\bibnamefont{Wokaun}},
  \emph{\bibinfo{title}{Principles of Nuclear Magnetic Resonance in One and Two
  Dimensions}} (\bibinfo{publisher}{World Scientific Publishing},
  \bibinfo{address}{New Jersey}, \bibinfo{year}{2004}).

\bibitem[{\citenamefont{Viola and Lloyd}(1998)}]{Viola1998}
\bibinfo{author}{\bibfnamefont{L.}~\bibnamefont{Viola}} \bibnamefont{and}
  \bibinfo{author}{\bibfnamefont{S.}~\bibnamefont{Lloyd}},
  \bibinfo{journal}{Phys. Rev. A} \textbf{\bibinfo{volume}{58}},
  \bibinfo{pages}{2733} (\bibinfo{year}{1998}).

\bibitem[{\citenamefont{Vandersypen and Chuang}(2004)}]{Vandersypen2004}
\bibinfo{author}{\bibfnamefont{L.~M.~K.} \bibnamefont{Vandersypen}}
  \bibnamefont{and} \bibinfo{author}{\bibfnamefont{I.~L.}
  \bibnamefont{Chuang}}, \bibinfo{journal}{Rev. Mod. Phys.}
  \textbf{\bibinfo{volume}{76}}, \bibinfo{pages}{1037} (\bibinfo{year}{2004}).

\bibitem[{\citenamefont{Cappellaro et~al.}(2007)\citenamefont{Cappellaro,
  Ramanathan, and G.Cory}}]{Cappellaro2007}
\bibinfo{author}{\bibfnamefont{P.}~\bibnamefont{Cappellaro}},
  \bibinfo{author}{\bibfnamefont{C.}~\bibnamefont{Ramanathan}},
  \bibnamefont{and} \bibinfo{author}{\bibfnamefont{D.}~\bibnamefont{G.Cory}},
  \bibinfo{journal}{Phys. Rev. Lett.} \textbf{\bibinfo{volume}{99}},
  \bibinfo{pages}{250506} (\bibinfo{year}{2007}).

\bibitem[{\citenamefont{Press et~al.}(2008)\citenamefont{Press, Ladd, Zhang,
  and Yamamoto}}]{Press2008}
\bibinfo{author}{\bibfnamefont{D.}~\bibnamefont{Press}},
  \bibinfo{author}{\bibfnamefont{T.~D.} \bibnamefont{Ladd}},
  \bibinfo{author}{\bibfnamefont{B.}~\bibnamefont{Zhang}}, \bibnamefont{and}
  \bibinfo{author}{\bibfnamefont{Y.}~\bibnamefont{Yamamoto}},
  \bibinfo{journal}{Nature} \textbf{\bibinfo{volume}{456}},
  \bibinfo{pages}{218} (\bibinfo{year}{2008}).

\bibitem[{\citenamefont{Phelps et~al.}(2009)\citenamefont{Phelps, Sweeney, Cox,
  and Wang}}]{Phelps2009}
\bibinfo{author}{\bibfnamefont{C.}~\bibnamefont{Phelps}},
  \bibinfo{author}{\bibfnamefont{T.}~\bibnamefont{Sweeney}},
  \bibinfo{author}{\bibfnamefont{R.~T.} \bibnamefont{Cox}}, \bibnamefont{and}
  \bibinfo{author}{\bibfnamefont{H.}~\bibnamefont{Wang}},
  \bibinfo{journal}{Phys. Rev. Lett.} \textbf{\bibinfo{volume}{102}},
  \bibinfo{pages}{237402} (\bibinfo{year}{2009}).

\bibitem[{\citenamefont{Mischuck et~al.}()\citenamefont{Mischuck, Deutsch, and
  Jessen}}]{Mischuck2009}
\bibinfo{author}{\bibfnamefont{B.}~\bibnamefont{Mischuck}},
  \bibinfo{author}{\bibfnamefont{I.~H.} \bibnamefont{Deutsch}},
  \bibnamefont{and} \bibinfo{author}{\bibfnamefont{P.~S.}
  \bibnamefont{Jessen}}, \bibinfo{note}{arXiv:0905.1094}.

\bibitem[{\citenamefont{Uhrig}(2007)}]{Uhrig2007}
\bibinfo{author}{\bibfnamefont{G.~S.} \bibnamefont{Uhrig}},
  \bibinfo{journal}{Phys. Rev. Lett.} \textbf{\bibinfo{volume}{98}},
  \bibinfo{pages}{100504} (\bibinfo{year}{2007}).

\bibitem[{\citenamefont{Uhrig}(2009)}]{Uhrig2009}
\bibinfo{author}{\bibfnamefont{G.~S.} \bibnamefont{Uhrig}},
  \bibinfo{journal}{Phys. Rev. Lett.} \textbf{\bibinfo{volume}{102}},
  \bibinfo{pages}{120502} (\bibinfo{year}{2009}).

\bibitem[{\citenamefont{Cloizeaux and Gaudin}(1966)}]{Cloizeaux1966}
\bibinfo{author}{\bibfnamefont{J.~D.} \bibnamefont{Cloizeaux}}
  \bibnamefont{and} \bibinfo{author}{\bibfnamefont{M.}~\bibnamefont{Gaudin}},
  \bibinfo{journal}{J. Math. Phys.} \textbf{\bibinfo{volume}{7}},
  \bibinfo{pages}{1384} (\bibinfo{year}{1966}).

\bibitem[{\citenamefont{Sutherland}(2005)}]{SutherlandBook}
\bibinfo{author}{\bibfnamefont{B.}~\bibnamefont{Sutherland}},
  \emph{\bibinfo{title}{Beautiful Models}} (\bibinfo{publisher}{World
  Scientific}, \bibinfo{address}{New Jersey}, \bibinfo{year}{2005}).

\bibitem[{\citenamefont{Diep}(1994)}]{DiepBook}
\bibinfo{author}{\bibfnamefont{H.~T.} \bibnamefont{Diep}},
  \emph{\bibinfo{title}{Frustrated Spin System}} (\bibinfo{publisher}{Oxford
  University Press}, \bibinfo{address}{Oxford}, \bibinfo{year}{1994}).

\bibitem[{\citenamefont{Haldane}(1982)}]{Haldane1982}
\bibinfo{author}{\bibfnamefont{F.~D.~M.} \bibnamefont{Haldane}},
  \bibinfo{journal}{Phys. Rev. B} \textbf{\bibinfo{volume}{25}},
  \bibinfo{pages}{4925} (\bibinfo{year}{1982}).

\bibitem[{\citenamefont{Mikeska and Kolezhuk}(2004)}]{Mikeska2004}
\bibinfo{author}{\bibfnamefont{H.-J.} \bibnamefont{Mikeska}} \bibnamefont{and}
  \bibinfo{author}{\bibfnamefont{A.~K.} \bibnamefont{Kolezhuk}},
  \bibinfo{journal}{Lect. Notes Phys.} \textbf{\bibinfo{volume}{645}},
  \bibinfo{pages}{1} (\bibinfo{year}{2004}).

\bibitem[{Bet()}]{Bethe}
\bibinfo{note}{H. A. Bethe, Z. Phys. {\bf 71}, 205 (1931); M. Karbach and G.
  M\"uller, Comput. Phys. {\bf 11}, 36 (1997).}

\bibitem[{\citenamefont{Avishai et~al.}(2002)\citenamefont{Avishai, Richert,
  and Berkovitz}}]{Avishai2002}
\bibinfo{author}{\bibfnamefont{Y.}~\bibnamefont{Avishai}},
  \bibinfo{author}{\bibfnamefont{J.}~\bibnamefont{Richert}}, \bibnamefont{and}
  \bibinfo{author}{\bibfnamefont{R.}~\bibnamefont{Berkovitz}},
  \bibinfo{journal}{Phys. Rev. B} \textbf{\bibinfo{volume}{66}},
  \bibinfo{pages}{052416} (\bibinfo{year}{2002}).

\bibitem[{\citenamefont{Santos}(2004)}]{Santos2004}
\bibinfo{author}{\bibfnamefont{L.~F.} \bibnamefont{Santos}},
  \bibinfo{journal}{J. Phys. A} \textbf{\bibinfo{volume}{37}},
  \bibinfo{pages}{4723} (\bibinfo{year}{2004}).

\bibitem[{\citenamefont{Hsu and d'Auriac}(1993)}]{Hsu1993}
\bibinfo{author}{\bibfnamefont{T.~C.} \bibnamefont{Hsu}} \bibnamefont{and}
  \bibinfo{author}{\bibfnamefont{J.~C.~A.} \bibnamefont{d'Auriac}},
  \bibinfo{journal}{Phys. Rev. B} \textbf{\bibinfo{volume}{47}},
  \bibinfo{pages}{14291} (\bibinfo{year}{1993}).

\bibitem[{\citenamefont{Kudo and Deguchi}(2005)}]{Kudo2005}
\bibinfo{author}{\bibfnamefont{K.}~\bibnamefont{Kudo}} \bibnamefont{and}
  \bibinfo{author}{\bibfnamefont{T.}~\bibnamefont{Deguchi}},
  \bibinfo{journal}{J. Phys. Soc. Jpn.} \textbf{\bibinfo{volume}{74}},
  \bibinfo{pages}{1992} (\bibinfo{year}{2005}).

\bibitem[{\citenamefont{Hahn}(1950)}]{Hahn-Echo}
\bibinfo{author}{\bibfnamefont{E.~L.} \bibnamefont{Hahn}},
  \bibinfo{journal}{Phys. Rev.} \textbf{\bibinfo{volume}{80}},
  \bibinfo{pages}{580} (\bibinfo{year}{1950}).

\bibitem[{\citenamefont{Carr and Purcell}(1954)}]{CP-Echo}
\bibinfo{author}{\bibfnamefont{H.~Y.} \bibnamefont{Carr}} \bibnamefont{and}
  \bibinfo{author}{\bibfnamefont{E.~M.} \bibnamefont{Purcell}},
  \bibinfo{journal}{Phys. Rev.} \textbf{\bibinfo{volume}{94}},
  \bibinfo{pages}{630} (\bibinfo{year}{1954}).

\bibitem[{\citenamefont{Viola and Knill}(2005)}]{Viola2005}
\bibinfo{author}{\bibfnamefont{L.}~\bibnamefont{Viola}} \bibnamefont{and}
  \bibinfo{author}{\bibfnamefont{E.}~\bibnamefont{Knill}},
  \bibinfo{journal}{Phys. Rev. Lett.} \textbf{\bibinfo{volume}{94}},
  \bibinfo{pages}{060502} (\bibinfo{year}{2005}).

\bibitem[{\citenamefont{Khodjasteh and Lidar}(2007)}]{Khodjasteh2005}
\bibinfo{author}{\bibfnamefont{K.}~\bibnamefont{Khodjasteh}} \bibnamefont{and}
  \bibinfo{author}{\bibfnamefont{D.~A.} \bibnamefont{Lidar}},
  \bibinfo{journal}{Phys. Rev. A} \textbf{\bibinfo{volume}{75}},
  \bibinfo{pages}{062310} (\bibinfo{year}{2007}).

\bibitem[{\citenamefont{Santos and Viola}(2008)}]{Santos2008}
\bibinfo{author}{\bibfnamefont{L.~F.} \bibnamefont{Santos}} \bibnamefont{and}
  \bibinfo{author}{\bibfnamefont{L.}~\bibnamefont{Viola}},
  \bibinfo{journal}{New J. Phys.} \textbf{\bibinfo{volume}{10}},
  \bibinfo{pages}{083009} (\bibinfo{year}{2008}).

\bibitem[{\citenamefont{Rego et~al.}(2009)\citenamefont{Rego, Santos, and
  Batista}}]{Rego2009}
\bibinfo{author}{\bibfnamefont{L.~G.~C.} \bibnamefont{Rego}},
  \bibinfo{author}{\bibfnamefont{L.~F.} \bibnamefont{Santos}},
  \bibnamefont{and} \bibinfo{author}{\bibfnamefont{V.~S.}
  \bibnamefont{Batista}}, \bibinfo{journal}{Annu. Rev. Phys. Chem.}
  \textbf{\bibinfo{volume}{60}}, \bibinfo{pages}{293} (\bibinfo{year}{2009}).

\bibitem[{not({\natexlab{a}})}]{noteCONVERGE}
\bibinfo{note}{The convergence of the expansions used to obtain the evolution
  operator in the presence of pulse sequences is discussed in
  Ref.~\cite{Santos2008} and references therein.}

\bibitem[{not({\natexlab{b}})}]{noteCHAOS}
\bibinfo{note}{For studies about what to expect for the dynamics of frustrated
  chaotic systems see, for instance,
  Refs.~\cite{rigol09,Santos2008PRE,Santos2009JMP,Langer2009,SantosPRE10}.}

\bibitem[{\citenamefont{Rabson et~al.}(2004)\citenamefont{Rabson, Narozhny, and
  Millis}}]{Rabson2004}
\bibinfo{author}{\bibfnamefont{D.~A.} \bibnamefont{Rabson}},
  \bibinfo{author}{\bibfnamefont{B.~N.} \bibnamefont{Narozhny}},
  \bibnamefont{and} \bibinfo{author}{\bibfnamefont{A.~J.}
  \bibnamefont{Millis}}, \bibinfo{journal}{Phys. Rev. B}
  \textbf{\bibinfo{volume}{69}}, \bibinfo{pages}{054403}
  (\bibinfo{year}{2004}).

\bibitem[{\citenamefont{Bowdrey et~al.}(2002)\citenamefont{Bowdrey, Oi, Short,
  Banaszek, and Jones.}}]{Bowdrey2002}
\bibinfo{author}{\bibfnamefont{M.~D.} \bibnamefont{Bowdrey}},
  \bibinfo{author}{\bibfnamefont{D.~K.~L.} \bibnamefont{Oi}},
  \bibinfo{author}{\bibfnamefont{A.~J.} \bibnamefont{Short}},
  \bibinfo{author}{\bibfnamefont{K.}~\bibnamefont{Banaszek}}, \bibnamefont{and}
  \bibinfo{author}{\bibfnamefont{J.~A.} \bibnamefont{Jones.}},
  \bibinfo{journal}{Phys. Lett. A} \textbf{\bibinfo{volume}{294}},
  \bibinfo{pages}{258} (\bibinfo{year}{2002}).

\bibitem[{not({\natexlab{c}})}]{noteBEST}
\bibinfo{note}{In this work we have not studied how particular initial states
  may affect our results. Our goal here was to show that control methods are
  viable techniques for the studies of the dynamics of quantum many-body
  systems. Studies of best-case performance may be found, for instance, in
  Ref.~\cite{Zhang2008} and references therein.}

\bibitem[{\citenamefont{Zhang et~al.}(2008)\citenamefont{Zhang, Konstantinidis,
  Dobrovitski, Harmon, Santos, and Viola}}]{Zhang2008}
\bibinfo{author}{\bibfnamefont{W.}~\bibnamefont{Zhang}},
  \bibinfo{author}{\bibfnamefont{N.~P.} \bibnamefont{Konstantinidis}},
  \bibinfo{author}{\bibfnamefont{V.~V.} \bibnamefont{Dobrovitski}},
  \bibinfo{author}{\bibfnamefont{B.~N.} \bibnamefont{Harmon}},
  \bibinfo{author}{\bibfnamefont{L.~F.} \bibnamefont{Santos}},
  \bibnamefont{and} \bibinfo{author}{\bibfnamefont{L.}~\bibnamefont{Viola}},
  \bibinfo{journal}{Phys. Rev. B} \textbf{\bibinfo{volume}{77}},
  \bibinfo{pages}{125336} (\bibinfo{year}{2008}).

\end{thebibliography}
\end{document}